\documentstyle[aps,prb]{revtex}
\begin{document}
\draft

\title{Coulomb Interaction and the Fermi Liquid State: Solution
by Bosonization}
\author{A. Houghton, H.-J. Kwon, J. B. Marston}
\address{Department of Physics, Brown University, Providence, RI 02912-1843}
\author{and R. Shankar}
\address{Sloane Laboratory of Physics, Yale University, New Haven, CT 06520}
\date{December 16, 1993}
\maketitle

\begin{abstract}
We investigate the effects of the Coulomb two-body interaction
on Fermi liquids via bosonization.  The Coulomb interaction is singular
in the limit of low momentum transfer, and recent
interest in the possibility that some singular interactions might destroy
the Fermi liquid state motivate us to reexamine it.
We calculate the exact boson correlation function to show that
the Fermi liquid state is retained in the case of Coulomb interactions. Spin
and charge degrees of freedom propagate together at the same velocity
and collective charge excitations (plasmons) exhibit the expected energy gap
in three dimensions.
Non-Fermi liquid behavior occurs, however, for a
super long-range interaction studied recently by Bares and Wen.

\end{abstract}

cond-mat/9312067

\section{Introduction}
The many body problem of fermions interacting via the Coulomb
interaction is central to condensed matter physics.  The standard
method for studying this problem is the summation of ring diagrams
which arise in perturbation theory: the random phase approximation
(RPA)\cite{Rick}.
Recently, another approach to this problem based on the renormalization group
(RG) has been developed\cite{Shankar}.
Both the RPA and the RG calculations assume that
the Fermi quasiparticle propagator retains the Fermi liquid form with
a simple pole.  This assumption is shown to be self-consistent since
in both approaches the bare Coulomb
interaction is screened down to a short-range form.
On the other hand, it is desirable to develop tools which do not
rely on the intermediate assumption of a Fermi liquid propagator to ascertain
whether non-Fermi liquid states are solutions of systems with singular
interactions.

Bosonization in dimensions greater than one\cite{Haldane,Tony,HKM}
is particularly
well suited to the study of singular long-range interactions as the realm in
which it is applicable, for low-energy excitations near the Fermi surface,
is precisely where these interactions are strongest.
Furthermore bosonization does not rely upon
a Fermi liquid form for the quasiparticle propagator; for example,
it encompasses the possibility of spin-charge separation.
Our results, however, are
unambiguous: we find that the Fermi liquid state is {\it the solution} to the
Coulomb problem.  By using the bosonization
transformation to determine the fermion quasiparticle
propagator, we obtain well-known results for the fermion
self energy: the imaginary
part is proportional to $\omega^2 \ln |\omega|$ in two dimensions and just
$\omega^2$ in three dimensions.  We emphasize that the bosonization method
yields non-perturbative information, so a natural next step would be to
use it to study the effects of transverse gauge interactions.

\section{Bosonization of the Two-Body Interaction}
We begin with the bare Hamiltonian for electrons or other fermions
interacting via the two-body
Coulomb or other long-range longitudinal interaction $V({\bf q})$:
\begin{equation}
H = \sum_{\bf k}~ \epsilon_{\bf k}~ c^{\dagger \alpha}_{\bf k}~
c_{{\bf k} \alpha} + {{1}\over{2 Vol}}~ \sum_{\bf k, p, q}~ V({\bf q})~
c^{\dagger \alpha}_{\bf k}~ c^{\dagger \beta}_{\bf p + q}~
c_{{\bf p} \beta}~ c_{{\bf k + q} \alpha}
\label{bare}
\end{equation}
where $\epsilon_{\bf k} \equiv {\bf k}^2/2m$ for the case of a Fermi gas
and there is an implicit sum
over repeated spin indices $\alpha$ and $\beta$.  Then we integrate out the
high-energy Fermi degrees of freedom with the use of the renormalization
group.\cite{Shankar,Others}  The resulting low-energy effective theory is
expressed in terms of quasiparticles $\psi_{{\bf k} \alpha}$ which obey
canonical anticommutation relations and which are
related to the bare electron operators via the wavefunction renormalization
factor $Z_{\bf k}$:
\begin{equation}
\psi_{{\bf k} \alpha} = Z_{\bf k}^{-1/2}~ c_{{\bf k} \alpha}
\label{z}
\end{equation}
for momenta $\bf k$ restricted to a narrow shell of thickness $\lambda$
around the Fermi surface:
$k_F - \lambda/2 < |{\bf k}| < k_F + \lambda/2$.  The resulting low-energy
Fermi liquid Hamiltonian is:
\begin{equation}
H_{FL} = \sum_{\bf k}~ \tilde{\epsilon}_{\bf k}~ \psi^{\dagger \alpha}_{\bf k}~
\psi_{{\bf k} \alpha} + {{1}\over{2 Vol}}~ \sum_{\bf S, T; q}~ V({\bf q})~
J({\bf S; -q})~ J({\bf T; q}) + \{{\rm regular~ terms}\}
\label{FL}
\end{equation}
Here $\tilde{\epsilon}_{\bf k} \equiv {\bf k}^2/2m^*$ incorporates the mass
renormalization.  The charge current in a given Fermi surface
patch $\bf S$ [where ${\bf S} \equiv (\theta, \phi)$ in three dimensions]
is defined by:
\begin{equation}
J({\bf S; q}) \equiv \sum_{\bf k} \theta({\bf S; k + q})~
\theta({\bf S; k})~ \{ \psi^{\dagger \alpha}_{\bf k + q}~
\psi_{\alpha {\bf k}} - \delta^3_{\bf q, 0}~ n_{\bf k} \}\ .
\label{curk}
\end{equation}
Here $\theta({\bf S; k}) = 1$ if ${\bf k}$ lies inside a squat box
of dimensions $\lambda \times \Lambda ^{D-1}$ centered
at $\bf S$ and equals zero
otherwise.  The regular terms in Eq. (\ref{FL}) do not diverge in the $\bf q
\rightarrow 0$ limit and consist of both large-$q$ exchange processes and
effective interactions
generated by the high-energy electrons that have been integrated out.
We assume in the following that no superconducting or charge or spin
density wave instabilities arise from the regular terms.

We now review the main aspects of bosonization and refer the reader to
two previous papers\cite{Tony,HKM} for more details.  In $D$-spatial
dimensions, the Fermi quasiparticle
fields of spin-$\sigma$, $\psi_\sigma$,
may be expressed\cite{Tony} in terms of the abelian boson fields
$\phi_\sigma$ as:
\begin{equation}
\psi_{\sigma}({\bf S; x}) = {1\over\sqrt{Vol}}~ \sqrt{{\Omega}\over{a}}
e^{i{\bf k}_{\bf S}{\bf \cdot x}}
\exp \{i{\sqrt{4\pi}\over \Omega} \phi_{\sigma}({\bf S; x})\}~ O({\bf S}),
\label{bosonization}
\end{equation}
where the dependence on time, $t$, is included implicitly in
the spatial coordinates ${\bf x}$ and $\bf S$ labels the patch
on the Fermi surface with momentum ${\bf k_S}$.  $Vol$ is the volume of the
system which equals $L^D$ in D-dimensions; the factor of $\sqrt{Vol}$
is introduced to keep the
fermion anticommutation relations canonical.
Both the $\psi$ and $\phi$ fields live inside the squat
box centered on $\bf S$ with height $\lambda$ in the radial (energy) direction
and area
$\Lambda^{D-1}$ along the Fermi surface.
These two scales must be small in the
following sense: $k_F >> \Lambda >> \lambda$.  We satisfy these limits
by setting $\lambda \equiv k_F/N$ and $\Lambda \equiv k_F/N^\alpha$ where
$0 < \alpha < 1$ and $N \rightarrow \infty$.
The quantity $a$ in the bosonization formula Eq. (\ref{bosonization})
is a real-space cutoff given by $a \equiv 1/\lambda$.  Here
$\Omega \equiv \Lambda^{D-1} (L/2 \pi)^D$ equals the number of states
in the squat
box divided by $\lambda$.  Finally, $O(S)$ is an ordering operator
introduced\cite{Tony,Luther}
to maintain Fermi statistics in the angular direction along
the Fermi surface.  (Anticommuting statistics are obeyed automatically
in the radial direction.)

With the connection Eq. (\ref{bosonization})
between the fermion and boson fields we may bosonize the free Hamiltonian.
The result is quadratic in the $\phi $ fields:
\begin{eqnarray}
H_0  &=& v_F~ \sum_{\bf S} \int d^Dx~ \psi^{\dag \alpha}({\bf S; x})~
\bigg{\{}{{\hat{\bf n}_{\bf S}
{\bf \cdot \nabla}}\over{\rm i}} - k_F\bigg{\}}~ \psi_\alpha ({\bf S; x})
 \nonumber \\
&=& {{2\pi~ v_F}\over{\Omega~ Vol}}~
\sum_{\bf S} \int d^Dx~ \{(\hat{\bf n}_{\bf S}
{\bf \cdot \nabla})~ \phi_\alpha ({\bf S; x})\}^2 ~ .
\end{eqnarray}
The total
bosonized Hamiltonian may be written as $H = H_c + H_s$, exhibiting the
factorization into charge and spin sectors.
The charge Hamiltonian is bilinear in the current
operators\cite{Tony} $J({\bf S; q})$:
\begin{equation}
H_c = {{1}\over{2}}~ \sum_{\bf S, T} \sum_{\bf q} V_c({\bf S ,T; q})~
J({\bf S ; -q})~ J({\bf T; q})~. \label{hamc}
\end{equation}
Long-range interactions $V({\bf q})$ are incorporated into
$V_c$ as matrix elements that couple currents in different patches:
\begin{equation}
V_c ({\bf S, T; q}) = {{1}\over{2}}~ \Omega ^{-1} v_F~ \delta^{D-1}_{\bf S, T}
+ {{1}\over{Vol}}~ V({\bf q})~ . \label{intc}
\end{equation}
The charge currents are related to the charge boson field $\phi \equiv
\phi_\uparrow + \phi_\downarrow$ by
$J({\bf S; x}) = \sqrt{4 \pi} {\bf \hat{n}_S \cdot \nabla} \phi({\bf S; x})$.
They obey the equal-time U(1) Kac-Moody relations\cite{Tony}:
\begin{equation}
[J({\bf S; q})~ ,~ J({\bf T; p})] = 2~ \Omega~ \delta^{D-1}_{\bf S, T}~
\delta^D_{\bf q + p, 0}~ {\bf q \cdot \hat{n}_S}\ .\label{kacc}
\end{equation}
Similarly, the spin-sector is described by a Hamiltonian which has a
free part:
\begin{equation}
H_s = {{1}\over{2}}~ \sum_{\bf S} \sum_{\bf q} {{v_F}\over{2 \Omega}}~
J_z({\bf S; -q})~  J_z({\bf S; q})\ \label{hams}
\end{equation}
and an interaction part that depends on the regular
Fermi liquid spin-spin coefficients $f_s$.
Here the abelian spin currents $J_z$ commute with the charge currents and are
expressed in terms of the spin boson field
$\phi_z \equiv \phi_\uparrow - \phi_\downarrow$ by
$J_z({\bf S; x}) = \sqrt{4 \pi} {\bf \hat{n}_S \cdot \nabla}
\phi_z({\bf S; x})$.
The interaction term has two parts: a term
that couples the z-component of the spin currents in different patches and
a term that couples the x- and y- components which has the form:
\begin{equation}
\sum_{\bf S, T}~ f_s({\bf S, T})~ \int d^Dx~
\cos \{ {{\sqrt{4 \pi}}\over{\Omega}}~
[\phi_z({\bf S; x}) - \phi_z({\bf T; x})] \}\ .
\end{equation}
We note that the coefficients $f_s$
are invariant under the renormalization group transformations despite the
fact that the above term resembles the form of the bosonized BCS
interaction\cite{HKM}.  The crucial difference is that here
the boson fields do not appear in pairs at opposite points of the
Fermi surface.
The $\beta$-function for the coefficients $f_s$ therefore equals zero.
For simplicity we set $f_s({\bf S, T}) = 0$ in the following.

\section{Quantized Bosons}
In this section we calculate the exact boson Green's function in the
charge sector.
To simplify the calculation we restrict our attention to the
case of spherical (circular in two dimensions) Fermi surfaces
and set the Fermi velocity equal to one.  None of these simplifications
is essential.
First we write the charge currents in terms of boson operators that satisfy
canonical commutation relations.  The choice:
\begin{equation}
J({\bf S; q}) = \sqrt{2 \Omega~ |{\bf \hat{n}_S \cdot q}|}~
[ a({\bf S; q})~ \theta({\bf \hat{n}_S \cdot q}) + a^\dagger({\bf S; -q})~
\theta(-{\bf \hat{n}_S \cdot q}) ] \label{canon}
\end{equation}
with
\begin{equation}
[ a({\bf S; q}),~ a^\dagger({\bf T; p}) ] = \delta^{D-1}_{\bf S, T}~
\delta^D_{\bf q, p} \ ,
\end{equation}
and $\theta(x) = 1$ if $x > 0$ and is zero otherwise,
satisfies the U(1) Kac-Moody commutation relation Eq. (\ref{kacc}).
For convenience we
denote $a({\bf S; q})$ and $a({\bf S; -q})$ by $a_R({\bf S; q})$ and
$a_L({\bf S; q})$, respectively the right and left moving fields.
The momentum-frequency space propagator
\begin{equation}
i G_i({\bf S; q}, \omega) = \langle a_i({\bf S; q}, \omega)~
a^\dagger_i({\bf S; q},
\omega) \rangle
\end{equation}
is related to the propagator of the $\phi$ fields by:
\begin{equation}
\langle \phi_i({\bf S; q}, \omega)~ \phi_i({\bf S; -q}, -\omega) \rangle
= {{\Omega}\over{4 \pi {\bf \hat{n}_S \cdot q}}}~
\langle a_i({\bf S; q}, \omega)~ a^\dagger_i({\bf S; q}, \omega) \rangle\ .
\label{gphi}
\end{equation}
As the interaction $V({\bf q})$ has no dependence on the patch indices
$\bf S$ and $\bf T$, a straightforward generalization of our previous
calculation\cite{HKM} for the short-range interaction $F_0$ allows us to
construct the boson propagator.
The exact solution of the Dyson equation for the irreducible self
energy is given by:
\begin{equation}
\Sigma^I(S; {\bf q}, \omega) = {{2 \Lambda^{D-1}~
{\bf \hat{n}_S \cdot q}}\over{(2 \pi)^D}}~ {{V({\bf q})}\over{1 + V({\bf q})~
\chi^0({\bf q}, \omega)}} \ .
\label{self}
\end{equation}
The Lindhard function $\chi^0$ in two dimensions is given by:
\begin{eqnarray}
\chi^0(x) &=& N(0)~ \int_0^{2\pi} {{d\phi}\over{2 \pi}}~
{{\cos(\phi)}\over{\cos(\phi) - x - i \eta~ {\rm sgn}(\omega)}} \nonumber \\
&=& N(0)~ \bigg{\{} 1 - |x| {{\theta(x^2 - 1)}\over{\sqrt{x^2 - 1}}}
+ i |x| {{\theta(1 - x^2)}\over{\sqrt{1 - x^2}}} \bigg{\}}
\label{chi0} ~; ~ D = 2 \ ,
\end{eqnarray}
where $x \equiv {{\omega}\over{|{\bf q}|}}$ and $N(0)$ is the density of states
at the Fermi energy which in two dimensions equals $k_F/\pi$ in units where
$v_F = 1$.  In three dimensions the Lindhard function equals:
\begin{equation}
\chi^0(x) = N(0)~ \bigg{\{} 1 - {x \over 2 }\ln \big( {x+1 \over  x-1} \big)
\bigg{\}} ~;~ D = 3.
\end{equation}
Despite the fact that a perturbative expansion\cite{HKM} has been used as an
intermediate step to obtain Eq. (\ref{self}),
all terms in the expansion have been summed yielding
the exact non-perturbative result valid for arbitrary dimension $D$.
For example, when the interaction is short-ranged,
in $D = 1$ the resummed expansion yields the well known exact
result for a Luttinger liquid\cite{Tony}.

In the $N \rightarrow \infty$ limit in which ${\bf q} \rightarrow 0$ the
self energy simplifies because $V({\bf q}) \rightarrow \infty$.  For
finite-$x$ we obtain:
\begin{equation}
\Sigma^I({\bf S; q}, \omega) = {{2 \Lambda^{D-1} {\bf \hat{n}_S \cdot q}}
\over{(2 \pi)^D~ \chi^0(x)}}\ .
\label{selfsimp}
\end{equation}
Screening is apparent at finite-$x$,
even in the boson propagator, as the self energy in the
small-q limit no longer depends on $V({\bf q})$.
The velocity is renormalized slightly from its bare value of unity,
$v_F^\prime = 1 + O({{\Lambda}\over{k_F}})^{D-1}$, and
the boson lifetime is now finite because of scattering into different
patches.  Note, however, that these changes represent irrelevant corrections
as the self energy Eq. (\ref{self}) scales to zero as $\Lambda \rightarrow 0$.
In particular, the pole in the boson propagator remains unchanged as
$N \rightarrow \infty$.  In the opposite $\omega$-limit\cite{Neto}
of $x \rightarrow
\infty$ the denominator in Eq. (\ref{self}) vanishes at frequencies
corresponding to the eigenvalues of the collective mode equation; consequently
the self energy diverges at these frequencies and in the three dimensional
Coulomb case the plasmons acquire a gap as expected\cite{Rick}.

\section{Fermion Quasiparticle Properties}

In the previous section we saw that long-range longitudinal interactions
modify
the boson propagator.  Though BCS scattering processes were ignored,
small-angle scattering processes made the boson lifetime finite.
With these results we can use
the bosonization formula Eq. (\ref{bosonization}) to infer the fermion
quasiparticle self energy.  Since bosonization is
carried out in $({\bf x}, t)$ space we must carry out three operations.
First we Fourier transform the charge and spin boson propagators into real
space and use the abelian relation $\phi_{\uparrow,\downarrow} = {{1}\over{2}}
(\phi \pm \phi_z)$ to construct the boson propagator for, say, up spins.
Next the exponential of the resulting expression yields the
fermion propagator in real space.
Finally an inverse Fourier transform of the fermion propagator back
into momentum space allows us to extract the self energy.

It is difficult technically to perform these steps in all generality.
It will be sufficient for our purposes to estimate the imaginary
part of the fermion self energy.
The charge boson Green's function may be written in configuration space as:
\begin{eqnarray}
i~ G_\phi({\bf S; x}, t) \equiv
\langle \phi({\bf S; x}, t)~ \phi({\bf S; 0}, 0)~
-~ \phi^2({\bf S; 0}, 0) \rangle
&=& i~ FT~ \big{\{} G_\phi \big{\}}
\nonumber \\
&=& i~ FT~ \big{\{} G_\phi^{(0)} + G_\phi^{(1)} \big{\}}
\label{RI}
\end{eqnarray}
where $FT$ represents the Fourier transform operation that converts
the variables $({\bf q}, \omega)$ to $({\bf x}, t)$.
In the second line,
$i G_\phi$ which is given by Eq. (\ref{gphi}), has been expanded in powers
of the vanishingly small parameter $\Lambda/k_F$.
The Fourier transform of the real term,
$i~ FT \big{\{} G_\phi^{(0)} \big{\}}$, yields the free propagator:
\begin{eqnarray}
\langle \phi({\bf T; x})~ \phi({\bf T; 0})
- \phi^2({\bf T; 0}) \rangle
&=& {\Omega ^2 \over 4\pi }
\ln ({i a \over {\bf x\cdot \hat{n}}_T - t})~ ;\ |{\bf x_\perp}
\Lambda| << 1
\nonumber \\
&\rightarrow& -\infty~ ;\ |{\bf x_\perp} \Lambda| >> 1
\label{gphi0}
\end{eqnarray}
where $\bf x_\perp$ denotes spatial directions perpendicular to the surface
normal $\bf \hat{n}_T$.  The contribution to the
imaginary part of the fermion self energy comes from $G_\phi^{(1)}$
in the limit of $|x| < 1$ for which quasiparticle damping occurs:
\begin{eqnarray}
G_\phi^{(1)}({\bf S; q}, \omega)
&=& {{\Omega}\over{4\pi}}~
{{\Sigma^I({\bf S; q}, \omega)}\over{\bf \hat{n_S} \cdot q}}~
[\omega - {\bf \hat{n}_S \cdot q} + i \eta~ {\rm sgn}(\omega)]^{-2}
\nonumber \\
&=& {{\Omega}\over{4\pi}}~
{2 \Lambda ^{D-1}\over{(2\pi)^D}~ \chi^0({\bf q}, \omega)}~
[\omega - {\bf \hat{n}_S \cdot q} + i \eta~ {\rm sgn}(\omega)]^{-2}\
\label{gphi1}
\end{eqnarray}
where the second line holds only for finite-$x$ and small $\bf q$.
The imaginary part of the fermion self energy at the pole may now be estimated
by exponentiating Eq. (\ref{RI}) and expanding to first order in the imaginary
part.  Following the steps in a previous paper\cite{HKM} we obtain for $D = 2$:
\begin{equation}
{\rm Im}~ \Sigma_f^{(1)}(\omega)|_{\rm pole} \approx
{1\over{N(0)(2\pi)^2}}~
{\rm sgn}(\omega)~ \bigg{\{}
\omega^2~ \ln {{2 |\omega|}\over{\Lambda}} - {{\omega^2}\over{2}}
\bigg{\}}
\end{equation}
which form we recognize immediately from
previous work on the quasiparticle lifetime within the RPA
approximation\cite{Quinn}.
This quantity is always negative since $|\omega | << \Lambda $.
In three dimensions the calculation yields:
\begin{equation}
{\rm Im}~ \Sigma_f^{(1)}(\omega)|_{\rm pole} \propto
{{\Lambda}\over{k_F^2}}~ \omega^2\ + {\rm O}(\omega^3/k_F^2) ;
\end{equation}
the appearance of the momentum cutoff $\Lambda$ in this expression coincides
with that found in traditional Fermi liquid theory\cite{Baym}.
In the classic RPA calculation\cite{Ferrell} the cutoff $\Lambda$ is replaced
by an energy of order the plasma frequency due to a different weighting of
the high-energy states.

In the case of Coulomb interactions the weight of the
quasiparticle pole, $Z_F$, remains non-zero at the Fermi surface.
In contrast, super long-range interactions {\it can} destroy the quasiparticle
pole.  For example, Bares and Wen\cite{Wen} studied an interaction in two
dimensions with a logarithmic potential given in momentum space
by $V({\bf q}) = g / q^2$ and showed that $Z_F = 0$
within RPA.  Note that $g$ is a momentum scale of order the Fermi momentum
and the plasmon gap is
non-zero and equals $\sqrt{g k_F/2 \pi}$ due to the super long-range nature
of the interaction.
In fact, this system may form a Wigner crystal; we ignore
any such $2 k_F$ instabilities in the following analysis.
We confirm that the quasiparticle pole is destroyed
in the super long-range case via bosonization.
One might be tempted to compute $Z_F$ by using the Kramers-Kronig relation
to derive the real part of the self energy from the imaginary part
estimated above.  This procedure however, is unreliable, as
the Kramers-Kronig relations involve an
integral over the entire energy range whereas the calculation breaks down at
energies greater than $\lambda$.  Instead we directly calculate the
real-space Fermi two-point function.  To do this we compute
the Fourier transform of Eq. (\ref{gphi1}) in the $|x| > 1$ region for which
the self energy Eq. (\ref{self}) is purely real.  For simplicity we also
expand the Lindhard function in powers of $1/x$:
\begin{equation}
\chi^0(x) = -{{N(0)}\over{2 x^2}} + O (1/x^4)\ .
\end{equation}
The important point is that, in the case of the equal-time two-point function,
the integral over $\bf q$ diverges logarithmically
in the infrared due to a factor of $q^2$ appearing in the
denominator of the momentum integral. Setting $t = 0$ we obtain:
\begin{equation}
\ln G_f({\bf S; x}) \approx -\ln ({\bf \hat{n}_S \cdot x})
+ \int_{q_c < |{\bf q}| < \Lambda} {{d^2q}\over{(2 \pi)^2}}~
{{\exp(i {\bf q \cdot x})}\over{q^2}}~
\int^{\lambda}_{|{\bf q}|} {{d\omega}\over{\pi}}~
{{1}\over{1/g - k_F/(2 \pi \omega^2)}}~
{{1} \over{[\omega - {\bf \hat{n}_S \cdot q} + i \eta~ {\rm sgn}(\omega)]^2}}
\end{equation}
Since $\omega > |{\bf q}|$ in the integral, we make the approximation
of replacing
$[\omega - {\bf \hat{n}_S \cdot q} + i \eta~ {\rm sgn}(\omega)]^2$
with $\omega^2 + i \eta^\prime$ in the denominator and carry
out the integrations.  Upon exponentiating the boson Green's function
we obtain the Fermi quasiparticle propagator in configuration space:
\begin{equation}
G_f({\bf S; x}) \approx {{(q_c |{\bf x}|)^\zeta}
\over{\bf \hat{n}_S \cdot x}}~ e^{-i k_F {\bf \hat{n}_S \cdot x}}~
;\ |{\bf \hat{n}_S \cdot x} | \Lambda >> 1,\
|{x_\perp}| \Lambda << 1
\label{gf}
\end{equation}
where
\begin{equation}
\zeta = {{1}\over{\pi}}~ \sqrt{{{g}\over{2 \pi k_F}}}~
\tanh^{-1} \sqrt{{{2 \pi \lambda^2}\over{g k_F}}} > 0 \ .
\end{equation}
The appearance of the infrared cutoff $q_c = O(1/L)$ in the numerator
reflects the super long-range nature of the Bares and Wen interaction.
In a Luttinger liquid with short-range interactions,
a length scale of order the lattice cutoff would appear in place of
the system size $L$.
The appearance of the anomalous exponent $\zeta$ is a
consequence of bosonization which treats the interaction non-perturbatively
and thus improves upon RPA which gives instead a logarithmic dependence on the
system size\cite{Wen}.  Evidently, the quasiparticle pole has been
destroyed by the super long-range interaction.
An evaluation of the fermion propagator
in the opposite limit of $\bf x = 0$ and at large times $t$, however, yields
the usual $t^{-1}$ Fermi liquid form as screening {\it is}
effective at low frequencies.  Likewise, thermodynamic properties which
are defined in this equilibrium $q$-limit such as
the specific heat also have the usual Fermi liquid form.  Note that in either
limit the propagator is odd (apart from the phase factor)
under $({\bf x}, t) \rightarrow -({\bf x}, t)$ as demanded by Fermi statistics.

Repeating the calculation in the case of Coulomb interactions in either
two or three dimensions we easily find that the propagator has the
standard Fermi liquid form.

\section{Conclusion}
We examined the effects of long-range interactions on fermion liquids
by bosonization in dimensions greater than one.
We find that the Fermi liquid state is the only solution to the problem
of a degenerate gas of fermions interacting via the Coulomb
two-body interaction in two and three spatial dimensions.
Bosonization allows us to go beyond an assumed Fermi
liquid form for the quasiparticle propagator.  Indeed,
the fact that the bosonized Hamiltonian separates into charge and
spin parts, $H = H_c + H_s$, leads to the possibility that,
as in one dimension\cite{Tony},
the quasiparticle propagator might also
exhibit spin-charge separation, especially in the case of the Coulomb
interaction which is singular.
Spin-charge separation in dimensions larger than
one would, however, destroy the Fermi liquid
as the key element, the existence of a
pole in the single-particle Green's function with non-zero spectral weight,
would be replaced by a branch cut and the pole would be destroyed.
This does not happen because, as we saw at the end of section III,
the location of the pole of the boson propagator is unchanged from its
free value in the $\Lambda \rightarrow 0$ limit.  Consequently the spin and
charge velocities are equal and both degrees of freedom
propagate together in the usual quasiparticle form.  On the other hand,
the Fermi liquid form is destroyed in the case of the super long-range
interaction studied by Bares and Wen.

\section{Acknowledgements}
R. Shankar thanks N. Read for bringing Ref. [11] to his attention.
This research was supported by the National Science Foundation
through grants DMR-9008239 (A.H.), DMR-9357613 (J.B.M.) and DMR-9120525 (R.S.).

\end{document}